\documentclass[12pt]{article}
\usepackage{cite,epsfig,amssymb,amsmath,graphicx,color}

\allowdisplaybreaks

\evensidemargin \oddsidemargin

\newcommand{\nn}{\nonumber}

\begin{document}

\begin{titlepage}

\renewcommand{\thefootnote}{\fnsymbol{footnote}}


\vspace{15mm}
\baselineskip 9mm
\begin{center}
{\Large \bf Holographic Entanglement Entropy of Anisotropic Minimal Surfaces in LLM Geometries}
\end{center}

\baselineskip 6mm
\vspace{10mm}
\begin{center}
 Chanju Kim,$^1$ Kyung Kiu Kim,$^2$ O-Kab Kwon,$^3$
 \\[10mm]
  $^1${\sl Department of Physics, Ewha Womans University,
  Seoul 120-750, Korea}
  \\[3mm]
  $^2${\sl Department of Physics,
   College of Science, Yonsei University, Seoul 120-749, Korea}
   \\[3mm]
  $^3${\sl Department of Physics, BK21 physics Research Division, Institute of Basic Science,
  \\Sungkyunkwan University,
Suwon 440-746, Korea}
     \\[10mm]
  {\tt cjkim@ewha.ac.kr,~kimkyungkiu@gmail.com,~okab@skku.edu}
\end{center}

\thispagestyle{empty}

\vfill
\begin{center}
{\bf Abstract}
\end{center}
\noindent
We calculate the holographic entanglement entropy (HEE) of the $\mathbb{Z}_k$
orbifold of Lin-Lunin-Maldacena (LLM) geometries which 
are dual to the vacua of the mass-deformed ABJM theory with Chern-Simons level $k$. 
By solving the partial
differential equations analytically, we obtain the HEEs for all LLM 
solutions with arbitrary M2 charge and $k$ up to $\mu_0^2$-order where $\mu_0$ is
the mass parameter. The renormalized entanglement entropies
are all monotonically decreasing near the UV fixed point
in accordance with the $F$-theorem. Except the multiplication factor and
to all orders in $\mu_0$, they are
independent of the overall scaling of Young diagrams which characterize 
LLM geometries. Therefore we can classify the HEEs of LLM geometries 
with $\mathbb{Z}_k$ orbifold in terms of the shape of Young diagrams modulo 
overall size. HEE of each family is a pure number independent of 
the 't Hooft coupling constant except the overall multiplication factor. 
We extend our analysis to obtain HEE analytically to $\mu_0^4$-order for 
the symmetric droplet case.
\\ [13mm]


\end{titlepage}

\baselineskip 6.6mm
\renewcommand{\thefootnote}{\arabic{footnote}}
\setcounter{footnote}{0}


\newpage

\section{Introduction}

Gauge/gravity duality has been a central paradigm for decades in theoretical 
physics. Among others, holographic calculation of the entanglement entropy 
(EE)~\cite{Ryu:2006bv,Ryu:2006ef} draws recently much attention due to 
its elegance and implications for the nature of quantum field theories as well as quantum gravity.

In this paper, we consider $\mathbb{Z}_k$ orbifolds of 
Lin-Lunin-Maldacena(LLM) geometries~\cite{Bena:2004jw,Lin:2004nb} 
with SO(2,1)$\times$SO(4)$\times$SO(4) isometry in 11-dimensional supergravity 
and calculate the holographic entanglement entropy (HEE) to nontrivial
orders in the mass parameter.
The main motivation is their connection to the 
Aharony-Bergman-Jafferis-Maldacena (ABJM) theory with level $k$~\cite{Aharony:2008ug} 
which is a conformal field theory (CFT) describing the 
dynamics of  M2-branes on the transverse 
$\mathbb{C}^4/\mathbb{Z}_k$ orbifold with the Chern-Simons level $k$. 
It allows a mass-deformation~\cite{Hosomichi:2008jb,Gomis:2008vc} which
preserves full $\mathcal{N}=6$ supersymmetries.
This mass-deformed ABJM (mABJM) theory has many
supersymmetric vacua. It has been shown that the vacua have one-to-one
correspondence with the $\mathbb{Z}_k$ orbifold~\cite{Kim:2010mr,Cheon:2011gv} 
of LLM geometries, which are classified by a 1-dimensional droplet picture, or equivalently 
Young diagrams~\cite{Lin:2004nb}.

The LLM metric has a mass parameter $\mu_0$ 
which is proportional to the mass parameter $\mu$ in the mABJM theory. 
Then we can explore the renormalization group (RG) 
flow of the renormalized entanglement entropy (REE)~\cite{Liu:2012eea}
triggered by the mass deformation from the ABJM theory
as a UV fixed point~\cite{Kim:2014yca}. 
Since there are many vacua in the theory, the RG flow depends on 
the vacuum. This should be manifested in the holographic calculation of 
REE for LLM geometries.
See also \cite{Klebanov:2012va,Nishioka:2014kpa,Rosenhaus:2014nha,Casini:2014yca,
Park:2015dia,Jones:2015twa,Casini:2015ffa} for the behavior of EE under relevant perturbations from 
the UV fixed point. 

An important issue related to REE is about the $c$-theorem which
states that there exists a $c$-function which is positive definite and
monotonically decreasing along the RG flow
\cite{Zamolodchikov:1986gt,Cardy:1988cwa,Komargodski:2011vj}.
In 3-dimensions in particular, it is called the 
$F$-theorem \cite{Jafferis:2011zi} because 
the free energy on a three sphere plays the role of $c$-function. 
The $F$-theorem
was proved~\cite{Casini:2012ei} through the connection of the free energy 
with the constant term of EE of 
a circle~\cite{Myers:2010xs,Casini:2011kv,Dowker:2010yj}.
In this paper, we will examine explicitly how $F$-theorem is realized in 
the HEE of the mABJM theory which has a large number of discrete vacua.

The LLM geometries with $\mathbb{Z}_k$ orbifold are all asymptotic to 
AdS$_4\times S^7/\mathbb{Z}_k$. They are, however, not spherically 
symmetric along the radial direction of the AdS geometry but depend on 
two transverse coordinates. Therefore, it is not a simple exercise to get the
minimal area for a given entangling region because one has to solve a partial
differential equation (PDE) for the two transverse coordinates. In the
previous work~\cite{Kim:2014yca}, the angle dependence was neglected to
simplify the calculation with the assumption that it would not contribute
at least in the leading order in $\mu_0$. Though sensible results were
obtained for simple droplet configurations, there were cases that the 
$F$-theorem is violated in this approximation. 
In this work, however, we take into account
all the angle dependence exactly. In other words, we solve the PDE exactly
for all LLM solutions with $\mathbb{Z}_k$ orbifold and obtain the
corresponding HEE up to $\mu_0^2$-order. Then we verify that
the REE satisfies the $F$-theorem for all relevant deformations connected 
to dual LLM geometries with $\mathbb{Z}_k$ orbifold. 
For some simple
droplet configurations with general $k$, we further extend our analysis to 
$\mu_0^4$-order and obtain REE analytically.

Since we work with the most general $k$ and the rank of the gauge group $N$,
it is possible to investigate the dependence of EE on these parameters
including the 't Hooft coupling $\lambda=N/k$ in particular. 
Note, however, that it is 
not a trivial task to compare the EEs with different $N$ 
or $k$ because they will not uniquely specify a droplet due to
many degeneracies. That is, in the field theory language, different vacua
will give different EEs and to begin with one has to specify the vacua 
to compare. We will see that, depending on which vacua to choose, the EE 
depends on $\lambda$ differently. Moreover, we will show that, up to
a multiplication factor, the HEE of LLM geometries is independent
of the overall scaling of the droplet configurations to all orders in $\mu_0$.
Therefore, we can classify the LLM geometries with $\mathbb{Z}_k$ orbifold 
in terms of the shape of the corresponding Young diagrams modulo overall size. 
At each order in $\mu_0$, they
are pure numbers independent of $\lambda$. These can be considered
as nontrivial results to test the gauge/gravity duality in the large $N$ limit between the 
LLM geometry and mABJM theory which are not conformal. 

This paper is organized as follows. 
In section 2, we briefly review the relation between the vacua of the 
mABJM theory and the droplet classification of the LLM geometry with 
$\mathbb{Z}_k$ orbifold. 
In section 3, we solve the PDE exactly to obtain the HEE of a disk 
up to $\mu_0^2$-order. We show that the resulting REE 
satisfies the $F$-theorem near the UV fixed point for all LLM geometries with 
$\mathbb{Z}_k$ orbifold. Then we show that it is classified by the shape
of the Young diagrams and discuss how it depends on $N$ and $k$.
We also calculate the REE analytically up to $\mu_0^4$-order for
simple droplets.
We draw our conclusion in section 4.

\section{HEE of the mABJM Theory and LLM Geometries}

Supersymmetric vacua of the mABJM theory are classified by the occupation numbers 
$(N_n,\, N_n')$~\cite{Cheon:2011gv}, which are numbers of irreducible 
$n\times (n+1)$ and $(n+1)\times n$ GRVV matrices~\cite{Gomis:2008vc}, 
respectively.
On the other hand, the LLM solutions with $\mathbb{Z}_k$ orbifold are also 
classified by the discrete torsions $(l_n,\, l_n')$ assigned 
in the droplet picture of the LLM geometry. 
It was shown that there exists one-to-one correspondence between $(N_n,\, N_n')$ and 
$(l_n,\, l_n')$ in the range, $0\le N_n,N_n',l_n,l_n'\le k$~\cite{Cheon:2011gv}.
Since the mass deformation of the ABJM theory is a relevant deformation from 
the UV fixed point, the dual LLM geometry with  $\mathbb{Z}_k$ orbifold is 
asymptotic to AdS$_4\times S^7/\mathbb{Z}_k$. 
We investigate the behavior of the RG flow near the UV fixed point 
in terms of the HEE for all LLM solutions with general $k$ and examine the 
$F$-theorem. 

Let us start with the LLM geometry dual to the vacua of the
$U(N)_k \times U(N)_{-k}$ mABJM theory with a mass parameter $\mu$.
The metric is given by
\begin{align}\label{metric0}
ds^2 = |G_{tt}|(-dt^2 + dw_1^2+dw_2^2) +G_{xx}\left(dx^2 + dy^2 \right) + G_{\theta\theta} ds_{S^3/\mathbb{Z}_k}^2 + G_{\tilde\theta \tilde\theta} ds_{\tilde S^3/\mathbb{Z}_k}^2,
\end{align}
where $ds_{S^3/\mathbb{Z}_k}^2$ and $ds_{\tilde S^3/\mathbb{Z}_k}^2$ are metrics
for two $S^3/\mathbb{Z}_k$ spheres
and
\begin{align} \label{gttxx}
&G_{tt} = -\left(  \frac{ 4\mu_0^2 y \sqrt{\frac{1}{4} - z^2} }{  f^2  }   \right)^{2/3},\quad 
G_{xx}= \left( \frac{f\sqrt{\frac{1}{4}- z^2    }}{  2\mu_0 y^2}  \right)^{2/3},
\\ \label{gtttt}
&G_{\theta\theta}=\left(  \frac{f y \sqrt{\frac{1}{2} + z }}{2 \mu_0 ( \frac{1}{2}-z )}  \right)^{2/3},\quad G_{\tilde \theta \tilde \theta}= \left( \frac{f y \sqrt{\frac{1}{2} - z }  }{ 2\mu_0 \left( \frac{1}{2}+z \right) }  \right)^{2/3}~
\end{align}
with
\begin{align} \label{deff}
f = \sqrt{1 - 4 z^2 - 4 y^2 V^2}.
\end{align}
In the metric, the mass parameter $\mu_0$ is identified with that
of the mABJM theory through 
$\mu_0 = \mu/4$ \cite{Cheon:2011gv} in the convention of \cite{Kim:2014yca}.
The geometry is completely determined by functions $z$ and $V$,
\begin{align}\label{zV}
z(x,y) = \sum_{i=1}^{2N_B + 1}  \frac{(-1)^{i+1} (x-x_i)}{ 2\sqrt{ (x-x_i)^2 + y^2 }},
\qquad V(x,y)= \sum_{i=1}^{2N_B+1} \frac{(-1)^{i+1}}{2 \sqrt{ (x-x_i)^2 + y^2 }},
\end{align}
where  $x_i$'s denote the boundaries of black and white strips in the droplet
representation with $N_B$ being the number of black droplets. 
For details of the droplet picture with general $k$, see \cite{Cheon:2011gv}.
Due to the quantization condition of the four-form fluxes on 4-cycles ending
on the edges of black/white regions, it turns out that $x_i$'s are quantized as 
\begin{align}\label{qntflx}
\frac{(x_{i+1}-x_{i})}{2\pi l_{{\rm P}}^3 \mu_0} \in \mathbb{Z},
\end{align}
where $l_{{\rm P}}$ is the Planck length. Note that the quantization is proportional
to $\mu_0$. It introduces $\mu_0$ dependence to the metric in addition to
the explicit dependence appearing in \eqref{gttxx} and \eqref{gtttt}.

The metric \eqref{metric0} goes asymptotically to 
$AdS_4 \times S^7/\mathbb{Z}_k$ with radius $R$ given by
\begin{equation}
R = (32 \pi^2 k \tilde N)^{1/6} l_{{\rm P}},
\end{equation}
where\footnote{%
$\tilde N$ is the area of the Young diagram made of positions $\tilde x_i$
divided by $k$, and is equal to the rank $N$ of the gauge group in the field
theory up to the contribution of discrete torsions 
\cite{Bergman:2009zh,Aharony:2009fc}. See section 3 for an example.
In addition, there is a further
constant correction $-\frac1{24}(k-\frac1k)$. With these corrections,
$\tilde N$ should eventually be the Maxwell M2 charge \cite{Cheon:2011gv},
i.e.\ $\tilde N = N - \frac1{24}(k-\frac1k)$.}
\begin{align} \label{coC}
\tilde N &= \frac1{2k}(\tilde C_2 - \tilde C_1^2), \nn \\
\tilde C_p &= \sum _{i=1}^{2N_B + 1} (-1)^{i+1}
    \left( \frac{x_i}{2\pi l_{{\rm P}}^3\mu_0}  \right)^p
 \equiv \sum _{i=1}^{2N_B + 1} (-1)^{i+1} {\tilde x_i}^p.
\end{align}
For later convenience, we define normalized coefficients $C_p$ by
\begin{equation} \label{coC2}
C_p \equiv (k \tilde N)^{-p/2} \tilde C_p,
\end{equation}
which are invariant under an overall scaling of $x_i$'s.
Then $C_2 - C_1^2 = 2$.

Now, let's consider a 9 dimensional surface in this geometry. 
We denote coordinates of the surface by $\sigma_i$ with $i=1, \ldots, 9$ and 
represent the embedding function as $X^M (\sigma^i)$ where $M=0, \ldots, 10$.
Then, the 9 dimensional area of the  surface  becomes
\begin{align}\label{gammaA0-1}
\gamma_A = \int d^9\sigma\, \sqrt{\det g_{ij}}
         = \int d^9\sigma\, \sqrt{\det 
G_{MN}  \ \frac{\partial X^M\partial X^N}{\partial \sigma^i\partial\sigma^j} 
},
\end{align}
where $g_{ij}$ is the induced metric of the surface
and $G_{MN}$ is the $11$-dimensional metric in (\ref{metric0}). 
The HEE is defined by \cite{Ryu:2006bv,Ryu:2006ef}
\begin{align}
S_A = \frac{{\rm Min}(\gamma_A)}{4G_N},
\end{align}
where $G_N=(2\pi l_{{\rm P}})^9/(32 \pi^2)$ is the 11 dimensional Newton constant.
In the next section we would like to calculate $S_A$ for disk-type entangling
surfaces in the small $\mu_0$ limit.

\section{Anisotropic Minimal Surfaces and HEE}

The effect of small mass deformation on the HEE has been considered in 
\cite{Kim:2014yca} under the approximation that the minimal
surface is independent of the angle in polar coordinates introduced below.
Though this approximation gives reasonable results consistent
with $F$-theorem for simple droplet configurations, one cannot expect that it
would be valid in general because the spherical symmetry is obviously broken. 
Indeed, for some droplet configurations the REE calculated in 
\cite{Kim:2014yca} does not decrease monotonically along the RG flow, 
violating the $F$-theorem. 
For these LLM geometries, it was also found that the curvature scalars are not small 
at some transverse positions even in the large $N$ limit~\cite{Kim:2014yca,Hyun:2013sf}, 
which implies that the gauge/gravity duality for those geometries does not
work in this approximation.  
In this section we would like to investigate the effect of small mass
deformation without resorting to such an approximation. In other words,
we will treat the angular dependence of the minimal surface exactly.
It amounts to solving PDEs with two
variables up to some nontrivial order in $\mu_0$.

From now on, we take only disk type entangling surfaces into account.
We will work with polar coordinates $u$ and $\alpha$ defined by
\begin{align}\label{polar}
x = \frac{R^3}{4lu} \cos\alpha,\qquad y = \frac{R^3}{4lu}\sin\alpha,
\end{align}
where $l$ is the radius of the disk at the boundary.
The minimal surface is bounded by a disk in the $w_1$-$w_2$ plane 
at the boundary of AdS space ($u=0$). To describe such a configuration, 
we may consider the following embedding,
\begin{align}
w_1 =\rho l \cos \sigma_1,\quad w_2=\rho l \sin\sigma_1,\quad 
u = u(\rho,\phi),\quad \alpha = \alpha( \rho,\phi ).
\end{align}
Plugging this into (\ref{gammaA0-1}), we obtain the action,
\begin{align}\label{mindisk}
\gamma_{{\rm disk}} = \frac{\pi^{5}  R^9}{8 k} \int_0^1
d\rho \int_0^\pi d\phi\,  \frac{g \rho \sin^3 \alpha}{u^2} 
\sqrt{ \alpha'^2 + \frac{u'^2}{u^2}
        + g^2 (\dot \alpha u' - \alpha' \dot{u})^2 },
\end{align}
where $\dot{} = \frac{\partial}{\partial \rho}$ and 
$'{}= \frac{\partial }{\partial \phi}$. We have also introduced the function 
$g(u, \phi)$ defined by
\begin{equation}
	f(u,\phi) = 2 \mu_0 l u \sin \phi\,g(u,\phi).
\end{equation}
Note that all the mass-deformation effect in \eqref{mindisk} appears
only through the function $g$ which contains the information of the droplet position $x_i$'s. 
In the undeformed limit $\mu_0 \rightarrow 0$,
$x_i$'s go to zero due to the quantization condition \eqref{qntflx}. 
Then it is easy to see that $g$ goes to unity and hence
\eqref{mindisk} reduces to the minimal surface action for the 
undeformed ABJM theory \cite{Ryu:2006bv}.

By utilizing the residual gauge degree of freedom, we may choose $\alpha=\phi$. 
Then the equation of motion yields a PDE for $u=u(\rho,\phi)$, 
\begin{align}\label{eqndsk}
\frac{\partial}{\partial\rho}\left( 
 \frac{\rho g^3 \dot{u}\sin^3\phi}{u^2\sqrt{{\mathcal X}}} \right)
+\frac{\partial}{\partial\phi}
 \left( \frac{\rho g u'\sin^3\phi}{u^4\sqrt{{\mathcal X}}}\right)
-\frac{\partial}{\partial u}
 \left( \frac{\rho g \sin^3\phi}{u^2}\sqrt{{\mathcal X} }  \right)=0,
\end{align}
where 
$
{\mathcal X} = 1 + \frac{u'^2}{u^2} + g^2 \dot{u}^2.
$
\subsection{HEE up to ${\cal O} (\mu_0^2)$}

Now we are ready to consider the effect of small mass deformation.
From \eqref{deff} and \eqref{zV} we see that
$g$ can be expanded in powers of $\frac{x_i}{\sqrt{x^2 + y^2}} \sim \mu_0 l u$.
Furthermore, $u$ itself will depend on $\mu_0 l$. As a result we can write
\begin{align}\label{mu exp}
	&u \equiv \sum_{n=0}^{\infty} u_n(\rho,\phi) (\mu_0 l)^n = u_0(\rho)
	+ u_1(\rho,\phi) {\mu_0 l} + u_2(\rho,\phi)(\mu_0 l)^2 + \cdots,
\nn \\
&g(u,\phi) \equiv \sum_{n=0}^{\infty} g_n(\phi) (u \mu_0 l)^n   \nn\\
&~~~~~~~~~= 1 + g_1(\phi) u_0\mu_0 l
+ [ g_1(\phi) u_1 +g_2(\phi) u_0^2  ] (\mu_0 l)^2 +\cdots.
\end{align}
Since $z$ and $V$ consist of the generating function of the Legendre
polynomials, $g_i$'s can be written in terms of Legendre polynomials
\cite{Kim:2014yca}. Explicitly, a few lower terms are 
\begin{align}\label{giphi}
g_1(\phi) &= D_1 \cos  \phi , \nn \\
g_2(\phi) &= D_2+D_3 \cos (2 \phi ),
\end{align}
where $D_i$'s are constants depending on the droplet positions,
\begin{align}
D_1 &= \frac1{\sqrt2}(C_3 - C_1 C_2), 
\notag \\
	D_2 &= -\frac1{32}  \left[ 4C_3^2 - 4C_1(C_1^2+C_2)C_3
- C_2^2(C_1^2 - 5 C_2) + 9(C_1^2  - C_2)C_4 \right]  , 
\notag \\
D_3 &= -\frac1{32}  \left[ 4C_3^2 - 4C_1(3C_1^2-C_2)C_3
 + C_2^2(C_1^2 + 3 C_2) + 15(C_1^2  - C_2)C_4  \right]  ,
\label{coD}
\end{align}
and $C_p$'s are defined in \eqref{coC2}.
Given $D_i$'s, one can solve the equations of motion \eqref{eqndsk}
perturbatively with respect to $\mu_0$ to obtain the change of the minimal
surface. Note that we have to solve inhomogeneous PDEs of two variables 
$\rho$ and $\phi$ in the background of lower order configurations. 
There is, in general, no guarantee that explicit form of solutions can be 
obtained. Nevertheless, in this case, we are able to find
exact solutions up to the nonvanishing second orders in perturbation.

Let us start with the zeroth order equation of (\ref{eqndsk}) in $\mu_0$,
\begin{align}
\ddot{u}_0 + \frac{(2\rho + u_0 \dot{u}_0)(1 + \dot{u}_0^2)}{\rho u_0} =0.
\end{align}
This is nothing but the equation of motion for the conformal case, as it
should be.
Imposing the boundary conditions, $u(0) =1$ and 
$\dot{u}(0) = 0 $, one can find the well-known 
solution which is a geodesic in AdS space,
\begin{align}\label{ms CFT}
u_0 (\rho) = \sqrt{1 - \rho^2}.
\end{align}
This gives the minimal surface for ABJM theory without mass deformation.

If we plug this solution into (\ref{eqndsk}), then the first order equation
of motion reads
\begin{align} \label{1stordereq}
&\rho (1-\rho^2)^2 \ddot{u}_1 + \rho u''_1 +(1-\rho^2)(1 - 2 \rho^2)\dot{u}_1
	+ 3 \rho\cot\phi\, u'_1 - 2 \rho u_1
\nn \\
&~~-D_1 \rho(1-\rho^2) (5  - 3\rho^2) \cos\phi =0.
\end{align}
We have to solve the equation under the boundary conditions
$u_1(1,\phi)=\dot{u}_1(0,\phi)=0$ and $u'_1(\rho,0)={u}'_1(\rho,\pi)=0$, where
the latter comes from the regularity at $\phi=0$ and $\pi$.
This is an inhomogeneous linear PDE with explicit
dependence on the independent variables $\rho$ and $\phi$. 
It, however, admits a very simple solution 
\begin{align}\label{solu1}
u_1(\rho, \phi) = -\frac{D_1}2 \left(1-\rho^2\right) \cos \phi.
\end{align}

One can proceed to the second order in $\mu_0^2$. The equation of motion
at the second order becomes, after 
the solutions (\ref{ms CFT}) and (\ref{solu1}) plugged into (\ref{mu exp}),
\begin{align}
&\rho(1-\rho^2)^2 \ddot{u}_2 + (1-\rho^2)(1- 2 \rho^2)\dot{u}_2
+ \rho ( u_2'' + 3\cot \phi\, u_2') -2\rho u_2
\nn \\
&+\frac18\rho (1-\rho^2)^{3/2}\Big[D_1^2(27 - 26 \rho^2) -16 D_2 (3- 2\rho^2)
\nn \\
&+ (11 D_1^2 - 16 D_3)(3-2\rho^2)\cos(2\phi)\Big] = 0~,
\end{align}
with the boundary conditions
$u'_2(\rho,0)={u}'_2(\rho,\pi)=0$ and 
$u_2(1,\phi)=\dot{u}_2(0,\phi)=0$. This is even more 
complicated than the first order equation \eqref{1stordereq}. 
Remarkably, however, a fully analytic solution is still available,
\begin{align}\label{genu2}
u_2(\rho,\phi) 
&= -\frac1{6\sqrt{1-\rho^2}} ( D_1^2 + 20 D_2 -12 D_3 )
\log (1 + \sqrt{1-\rho^2}) \nn \\
&~~+\frac1{48}\left\{[8 + (9-13\rho^2)\sqrt{1-\rho^2}]D_1^2
+16[10 - (6-\rho^2)\sqrt{1-\rho^2}]D_2 \right. \nn \\
&\left. ~~-48(2 - \sqrt{1-\rho^2}) D_3 \right\}
+ \frac1{48}(11D_1^2 - 16 D_3)(1-\rho^2)^{3/2}\cos (2\phi).
\end{align}

Having found the solution to the $\mu_0^2$-order, now we can calculate 
the minimal surface area (\ref{mindisk}) to this order,
\begin{align}
\gamma_{{\rm disk}} = \gamma_{{\rm disk}}^{(0)}+  \mu_0 l \gamma_{{\rm disk}}^{(1)}
+ (\mu_0 l)^2\gamma_{{\rm disk}}^{(2)}+\cdots.
\end{align}
Inserting the solutions (\ref{ms CFT}), (\ref{solu1}) and (\ref{genu2}) into 
(\ref{mindisk}), we obtain%
\footnote{
Introducing the UV cutoff $\epsilon$ in the $u$ coordinate changes the 
upper limit of the integration range of $\rho$ in \eqref{mindisk}.
}:
\begin{align}\label{gamma2disk}
\gamma_{{\rm disk}}^{(0)} =\frac{\pi^5 R^9}{6 k} \left( \frac{l}{\epsilon}-1 \right),~~
\gamma_{{\rm disk}}^{(1)} =0,~~
\gamma_{{\rm disk}}^{(2)} =-\frac{ \pi^5 R^9}{72 k} (12 D_3-D_1^2-20 D_2),
\end{align}
where $\epsilon$ in $\gamma_{disk}^{(0)}$ is the UV cutoff in the $u$
coordinate. Note that the first order correction vanishes due to the 
angular integration. For the second order contribution $\gamma_{disk}^{(2)}$,
it is crucial to notice that the combination 
$(12D_3-D_1^2-20D_2)$ 
can be rewritten in the form of a complete square,
\begin{align}
12D_3-D_1^2-20D_2
=16+ \frac12(C_3 - 3 C_1 C_2 +2 C_1^3)^2 ,
\end{align}
where we used the parameter relations in \eqref{coD}. 
The entanglement entropy then becomes
\begin{align}\label{HEEfinal}
S_{{\rm disk}}=\frac{\pi^5 R^9}{24 G_N k} \left\{ \frac{l}{\epsilon}-1 
	-\mu_0^2 l^2 \left[\frac43+ \frac1{24}(C_3 - 3C_1 C_2 +2 C_1^3)^2
                            \right]\right\} + \mathcal{O}(\mu_0^3).
\end{align}
A few comments are in order. First, it is not difficult to show that
the expression $(C_3 - 3C_1 C_2 +2 C_1^3)$
appearing here is a unique combination made of cubic terms
in $x_i$ which is invariant under the translation $x_i \rightarrow x_i + a$.
It provides a nontrivial consistency check of the result. 
Moreover, the expression appears in \eqref{HEEfinal} in the form
of a complete square and hence the second order term is negative definite. 
This has an important implication in relation to the $F$-theorem~%
\cite{Jafferis:2011zi, Myers:2010xs, Casini:2012ei}. 
In the present context, the REE can play the role of
a $c$-function of the theory \cite{Liu:2012eea}.
It is computed by the prescription 
\begin{align}\label{REEdef}
	{\cal F}_{{\rm disk}} &\equiv \left(l\frac{\partial}{\partial l} -1
\right)S_{{\rm disk}} \nonumber \\
&=\frac{\pi^5 R^9}{24 G_N k} 
\left\{ 1 -  \mu_0^2 l^2 \left[\frac43+ \frac1{24}(C_3 - 3C_1 C_2 +2 C_1^3)^2
\right] \right\}   + \mathcal{O}(\mu_0^3),
\end{align}
which is clearly monotonically decreasing near the UV fixed point. Note that this is true 
for {\em all} geometries dual to the vacua of mABJM, regardless of the
$C_i$'s. 
This result may be considered as an evidence of the validity of 
holography for non-conformal case. 
This corrects the result in \cite{Kim:2014yca}
where REE was calculated with the angle ($\alpha$) dependence neglected and
showed increasing behavior for some asymmetric droplet configurations. 
This means that the angle dependence of the minimal surface in the LLM geometry 
results in nontrivial contributions.

As is evident from the calculation, the REE depends on $N$ or $k$ only 
through $C_p$'s except the overall multiplication. Therefore, different
theories with same $C_p$'s will give essentially the same REE. 
Since $C_p$'s defined in \eqref{coC2} are invariant under the overall 
scaling of droplet boundaries $x_i$'s, a family of droplets with a same 
geometric shape of Young diagrams modulo overall size give the same 
REE up to a multiplication factor. 
This holds to all orders in $\mu_0$ since the action \eqref{mindisk} 
is completely determined in terms of $C_p$'s up to an overall factor. 

\begin{figure}
\begin{center}
\includegraphics[width=9cm,clip]{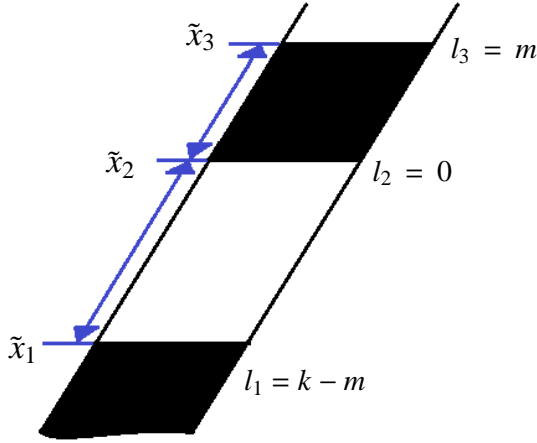}
\end{center}
\caption{
\label{fig:droplet}
A droplet representation of $N_B$=1 case.
$l_i$'s are discrete torsions assigned at $\tilde x_i$'s. 
}
\end{figure}

As a simple example, we consider the case of $N_B = 1$ with arbitrary $k$
for which the geometry is specified by three boundary positions 
$x_1$, $x_2$ and $x_3$ of the droplet. See Fig.\ref{fig:droplet}.
From \eqref{coC}, the REE becomes
\begin{align}\label{REEdef2}
{\cal F}_{{\rm disk}} =\frac{\pi^5 R^9}{24 G_N k} 
\left\{ 1 -  \mu_0^2 l^2 \left[\frac7{12}
 + \frac38 \left(\frac{x_3-x_2}{x_2-x_1} + \frac{x_2-x_1}{x_3-x_2} \right)
 \right] \right\} + \mathcal{O}(\mu_0^3).
\end{align}
This result explicitly shows that scaling the overall size of the droplet 
(or the shape of the Young diagram) does not change REE except the overall 
multiplication factor. 

To connect the result with the field theory,
let us parameterize the integer-valued positions 
$\tilde x_i = x_i/2\pi l_P^3 \mu_0$ defined in \eqref{coC} as
$\tilde x_1 = -pk -m$, $\tilde x_2 = (q-p)k$, $\tilde x_3 = qk +m$, 
where $p,q,m$ are positive integers and $0\le m < k$, 
so that the location of the Fermi level of the droplet is zero. 
Then including the contribution of the discrete torsions\cite{Cheon:2011gv},
we obtain the rank $N$ of the gauge group as 
$N = (p k +m)(qk +m)/k - m(m-k)/k = k pq + m(p+q+1)$. 
In the large $N$ limit with finite 't Hooft coupling $\lambda = N/k$,  
\eqref{REEdef2} is reduced to 
\begin{align}\label{REEdef3}
{\cal F}_{{\rm disk}} =\frac{\pi^5 R^9}{24 G_N k} 
\left\{ 1 -  \mu_0^2 l^2 \left[\frac7{12}
		+ \frac38 \left( \frac{p}q + \frac{q}p \right)
 \right] + \mathcal{O}(1/k) \right\} + \mathcal{O}(\mu_0^3),
\end{align}
where we assumed $k\gg m$ for simplicity and $\lambda = pq$ in this limit.
Therefore, for a fixed ratio of $p$ and $q$ the mass correction is
independent of $\lambda$. On the other hand, if we scale, say, $p$ with 
$q$ fixed, then the correction depends nontrivially on $\lambda$ since
$p/q = \lambda/q^2$. 
Note that by changing $p$ or $q$, we are comparing theories with
different $\lambda$. The above result demonstrates that 
$\lambda$-dependence of REE appears in different ways, depending on how vacua
are selected in mABJM theories for comparison.

Finally, let us give a comment on the stationarity of the RG flow.
Since the deformation parameter $\mu$ enters into
the mABJM theory as a mass of a supermultiplet, the deformation is of
the first order in $\mu$ due to the fermionic mass term. 
On the other hand, REE $\mathcal{F}$ in \eqref{REEdef} has vanishing 
first order correction in $\mu_0$. It means that
the RG flow at UV point is stationary. This is consistent with the
result of \cite{Nishioka:2014kpa}.

\subsection{HEE up to ${\cal O} (\mu_0^4)$ for symmetric droplet case}

For some simple droplet configurations, it is possible to obtain higher order 
corrections analytically. For example, if $D_1=0$, the first order
equation \eqref{1stordereq} becomes homogeneous with vanishing boundary 
conditions and hence the first order correction $u_1$ is zero identically.
This will also simplify higher order equations.  Here we will consider
a symmetric droplet case obtained by putting $p=q$ in Fig.\ref{fig:droplet},
i.e., $(\tilde x_1,\tilde x_2,\tilde x_3)=(-pk -m, 0, pk +m)$ and
$N = kp^2 + m(2p+1)$.

To consider higher order corrections up to $\mu_0^4$-order,
we need the following coefficient functions in (\ref{mu exp}),
\begin{align}\label{g46}
g_2(\phi) &= -\frac18 + \frac98\cos (2\phi), 
\nn \\
g_4(\phi) &= \frac{1}{256}
[85 -60 \cos (2 \phi )+359 \cos (4 \phi )],
\end{align}
as well as $g_{n}(\phi) =0$ for odd $n$.
By symmetry we may set $u_n=0$ for odd $n$.
Then employing all the lower order results including \eqref{ms CFT} and
\eqref{genu2} gives us the equation of motion for 
$u_4(\rho,\phi)$,
\begin{align}\label{eom u4}
\rho(1-\rho^2)^2 \ddot { u}_4 
+ (1-\rho^2)(1- 2 \rho^2)\dot {  u}_4
+ \rho ( {  u}_4'' + 3\cot \phi\, {  u}_4') -2\rho {  u}_4&
\nn \\
+B_0^{(4)}(\rho)+ B_2^{(4)}(\rho) \cos(2\phi) + B_4^{(4)}(\rho)\cos(4\phi) = 0&,
\end{align}
where 
\begin{align}\label{B4s}
B_0^{(4)}(\rho) &= \frac{64 \rho (3 \rho ^2+2)}{9 (1-\rho ^2)^{3/2}}
\left[\log\left(\sqrt{1-\rho ^2}+1\right)\right]^2
-\frac{2 }{9 \rho }(9 \rho ^6+28 \rho ^4+187 \rho ^2-64)
\nn \\
&-\frac{2 \rho  }{9 (1-\rho ^2)}\left[128 \rho ^2-(67 \rho ^4+8 \rho ^2+85)
\sqrt{1-\rho ^2}+192\right] \log \left(\sqrt{1-\rho ^2}+1\right)
\nn \\
&+\frac{1}{1152 \rho  \sqrt{1-\rho ^2}}
(6147 \rho^{10}-15629 \rho^8+6165 \rho^6-8463 \rho^4+69124 \rho^2-16384),\nn \\
B_2^{(4)}(\rho) &= \frac{3}{32} \rho  (1-\rho^2)
\left[576-192 \rho^2-(15 \rho^4-227 \rho^2+404) \sqrt{1-\rho ^2}\right]
\nn \\
&~~-6 \rho  (7-\rho ^2) \sqrt{1-\rho ^2} \log \left(\sqrt{1-\rho ^2}+1\right),
\nn \\
B_4^{(4)}(\rho) &= -\frac{125}{128} \rho (1-\rho^2)^{5/2} (4-3 \rho^2).
\end{align}
This is again an inhomogeneous
PDE with very complicated source terms but fortunately we can find the solution satisfying the necessary
boundary conditions,
\begin{align}\label{u4}
u_4(\rho, \phi) &=C_4^{(4)}(\rho)\cos(4\phi)
+C_2^{(4)}(\rho)\cos (2\phi) + C_0^{(4)}(\rho),
\end{align}
where
\begin{align}
C_4^{(4)}(\rho) &= - \frac{25}{256}  \left(1-\rho ^2\right)^{5/2},
\nn \\
C_2^{(4)}(\rho) &= \frac{1}{64 \left(1-\rho ^2\right)^{3/2}}\Big[  96 \left(7-2 \rho ^2\right)
\rho ^2 \log \left(\sqrt{1-\rho ^2}+1\right)+32
\left(6 \rho ^4-16 \rho ^2-5\right) \sqrt{1-\rho ^2}
\nn \\
&\hskip 3.3cm +\left(1-\rho ^2\right)
\left(3 \rho ^6-97 \rho ^4+209 \rho ^2+125\right)\Big],
\nn \\
C_0^{(4)}(\rho) &= -\frac{1}{\sqrt{1-\rho ^2}} \int_\rho^{1} F(s) \, ds,
\label{C04}
\end{align}
with
\begin{align}
F(s)&= \frac{135 s^6+598 s^4-3371 s^2+1423}{135 s (1-s^2)^{3/2}}
      -\frac{256}{9s} \sqrt{1-s^2} \log 2 \nn \\
     &+\frac{18441 s^8-37339 s^6-170949
       s^4+326871 s^2-182144}{17280 s (1-s^2)}
     -\frac{64 s \log ^2(\sqrt{1-s^2}+1)}{9 (1-s^2)^2} \nn \\
     &+\frac1{9s}\left[\frac{64 (s^4+1)}{(1-s^2)^{3/2}}
      -\frac{134 s^6 - 524 s^4 + 501 s^2 - 192}{(1-s^2)^2}\right]
        \log (\sqrt{1-s^2}+1).
\end{align}
Inserting the solutions in \eqref{ms CFT}, \eqref{genu2}, and \eqref{u4} 
into \eqref{mu exp} and \eqref{mindisk}, we easily obtain the corresponding HEE and REE 
for the symmetric droplet up to 
$\mu_0^4$-order,
\begin{align}\label{HREE}
S_{{\rm disk}}&=\frac{\pi ^5 R^9}{24 k G_N}\left( \frac{l}{\epsilon }-1
-\frac{4 }{3}\,\mu_0^2l^2+\frac{ 2671-3840 \log 2}{540}\, \mu_0^4l^4\right)
+ \mathcal{O}(\mu_0^6),
\nn \\
{\cal F}_{{\rm disk}} &= \frac{\pi ^5 R^9}{24 k G_N}\left( 1-\frac{4}{3} \mu_0^2l^2
+\frac{ 2671-3840 \log 2 }{180}\,\mu_0^4l^4\right)
+ \mathcal{O}(\mu_0^6). 
\end{align}
This result holds for arbitrary $N$ and $k$. In general, the expressions 
of the HEE and REE depend on $N$, $k$,  radius of the disk, 
and the choice of droplets as we have seen in the previous subsection.
However, for a family of droplets related by rescaling of the overall
size, the coefficients of the 
corrections are given by pure numbers as seen in \eqref{HREE}, and 
in particular are independent of the 't Hooft coupling constant $\lambda$.
This is an interesting phenomena from the point of view of the 
gauge/gravity duality. Based on the duality relation between the 
vacua of the mABJM theory and the LLM geometries with $\mathbb{Z}_k$ orbifold, 
one can examine the HEE conjecture by computing the EE of the dual field theory 
on a family of vacua considered here.

\section{Conclusion}

We investigated the RG flow behavior and the $F$-theorem in terms of the HEE 
near the UV fixed point in 3-dimensions,  where a supersymmetric Chern-Simons matter 
theory is living. As the UV CFT we considered the ${\cal N}=6$ ABJM theory 
and introduced the supersymmetry preserving mass deformation, called the mABJM theory. 
This deformation is a relevant deformation and so triggers the RG flow from the 
UV fixed point. 
To describe the RG behavior near the UV fixed point, we adapted the HEE conjecture 
to the LLM geometry in 11-dimensional supergravity, since 
the supersymmetric vacua of the mABJM theory 
have one-to-one correspondence with the LLM geometries with $\mathbb{Z}_k$ orbifold.  

The LLM solution has SO(2,1)$\times$SO(4)$\times$SO(4) isometry and warp factors 
of the metric depend on the two transverse coordinates $(u,\alpha)$ in \eqref{polar}. 
For this reason, one has to solve the PDE for $u$ and $\alpha$ to obtain 
the minimal surface in the HEE proposal. 
In this paper, we analytically solved the PDE up to $\mu_0^2$-order for all LLM solutions 
with arbitrary $N$ and $k$. 
We found that REEs have different RG trajectories depending on the 
LLM geometries but they are always monotonically decreasing near the UV fixed
point in accordance with the $F$-theorem 
in 3-dimensions.\footnote{Recently, it was reported that the 
strong form of the $F$-theorem, 
which describes the monotonically decreasing behavior of the REE, is violated 
for relevant deformations by operators of conformal dimension 
$3/2 < \Delta<5/2$~\cite{Taylor:2016aoi,Taylor:2016kic}. One needs 
more investigations in this direction.}
We also computed the REE up to $\mu_0^4$-order for a special family of
LLM geometries with arbitrary $N$ and $k$. It would be 
interesting to extend the result to more general case.

Since the HEE proposal is based on the gauge/gravity duality, in order to
compare our results in gravity side with those in the mABJM theory, one has to 
consider the large 't~Hooft coupling $\lambda=N/k$ in the $N\to\infty$ limit. 
In general, the effect of mass deformation in REE would depend on $\lambda$
and calculation in field theory side is a formidable task due to 
nonperturbative effects coming from the strong coupling constant. 
In the mABJM theory, there are further complications from the presence
of many vacua. However, we found that for a family of droplets with 
a same shape of Young diagrams, we have the same REE (or HEE) up 
to overall dependence on $N$ and $k$.
One might be able to calculate the REE in the field 
theory side perturbatively 
on a certain class of vacua, and compare 
the result quantitatively with that in the dual gravity side.

\section*{Acknowledgements}

KK thanks  Tatsuma Nishioka  for helpful comments in the 9th Asian Winter School on Strings in Busan.
This work was supported
by the National Research Foundation of Korea(NRF) grant with the grant number
NRF-2015R1D1A1A01058220 (K.K.) and NRF-2014R1A1A2059761 (O.K.). 

\end{document}